\documentclass[lettersize,journal]{IEEEtran}
\usepackage{amsmath,amsfonts}
\usepackage{array}
\usepackage[caption=false,font=normalsize,labelfont=sf,textfont=sf]{subfig}
\usepackage{textcomp}
\usepackage{stfloats}
\usepackage{url}
\usepackage{verbatim}
\usepackage{graphicx}
\usepackage{cite}
\usepackage{enumitem}
\usepackage{fourier}
\usepackage{circledsteps}
\pgfkeys{/csteps/fill color=green!10}
\usepackage{flushend}

\usepackage{subfig}
\usepackage{caption}

\usepackage[ruled,vlined,linesnumbered]{algorithm2e}

\SetCommentSty{mycommfont}

\DeclareMathAlphabet{\mathcal}{OMS}{cmsy}{m}{n}
\SetMathAlphabet{\mathcal}{bold}{OMS}{cmsy}{b}{n}

\newcommand{\var}[1]{\textrm{Var}{\left(#1\right)}}

\newcommand{\mo}{\textrm{~mod~}}
\def \ZZ {{\mathbb Z}}

\begin{document}

\title{Make Shuffling Great Again: A Side-Channel Resistant Fisher-Yates Algorithm for Protecting Neural Networks}
\author{Leonard Pu\v{s}k\'a\v{c}, Marek Benovič, Jakub Breier, and Xiaolu Hou
\thanks{
L. Pu\v{s}k\'a\v{c}, Marek Benovič and X. Hou are with Slovak University of Technology, Bratislava, Slovakia, e-mail: xpuskac@stuba.sk, marek.benovic@stuba.sk, houxiaolu.email@gmail.com

J. Breier is with TTControl GmbH, Vienna, Austria, e-mail: jbreier@jbreier.com}}



\maketitle

\begin{abstract}
Neural network models implemented in embedded devices have been shown to be susceptible to side-channel attacks (SCAs), allowing recovery of proprietary model parameters, such as weights and biases.
There are already available countermeasure methods currently used for protecting cryptographic implementations that can be tailored to protect embedded neural network models.
Shuffling, a hiding-based countermeasure that randomly shuffles the order of computations, was shown to be vulnerable to SCA when the Fisher-Yates algorithm is used.

In this paper, we propose a design of an SCA-secure version of the Fisher-Yates algorithm.
By integrating the masking technique for modular reduction and Blakely's method for modular multiplication, we effectively remove the vulnerability in the division operation that led to side-channel leakage in the original version of the algorithm.
We experimentally evaluate that the countermeasure is effective against SCA by implementing a correlation power analysis attack on an embedded neural network model implemented on ARM Cortex-M4.
Compared to the original proposal, the memory overhead is $2\times$ the biggest layer of the network, while the time overhead varies from $4\%$ to $0.49\%$ for a layer with $100$ and $1000$ neurons, respectively.
\end{abstract}


\section{Introduction}
\IEEEPARstart{N}{eural} network (NN) implementations have become increasingly deployed in embedded devices, being used for various applications from autonomous vehicles to smart IoT devices.
While these deployments offer great computational efficiency and real-time performance, they also introduce vulnerability to hardware-based attacks such as side-channel attacks (SCAs)~\cite{kocher1999differential}. 
SCAs exploit physical emanations, such as power consumption, electromagnetic leaks, or timing variations to extract the sensitive values used during the computation~\cite{hou2024cryptography}. 
Such physical emanations, commonly referred to as \textit{leakages}, can lead to the compromise of intellectual property by revealing model parameters -- weights and biases~\cite{batina2019csi}.

As the area of hardware security has been researched for decades in the field of cryptography, it is natural to assess which protection techniques can be adapted for NN implementations.
Masking, while providing strong security guarantees, incurs significant overhead when applied to the whole network~\cite{dubey2022modulonet}.
The number of parameters in NNs, ranging to millions, naturally calls for lightweight countermeasures that do not significantly increase the required memory or execution time.
This is especially true for embedded applications running on resource-constrained computing units.

Hiding-based countermeasures, such as shuffling, on the other hand, can greatly increase the attacker's effort while keeping the overhead manageable even in embedded systems~\cite{hou2024cryptography}.
Shuffling has been extensively evaluated by the cryptography community, for example, \cite{veyrat2012shuffling} provides an analysis of its application.
Herbst et al.~\cite{herbst2006aes} proposed it for shuffling Sboxes and key additions in a hardware AES implementation. 
The most recent paper by Xu et al.~\cite{xu2025hardware} modifies the Fisher-Yates algorithm for hardware usage to show no TVLA leakage with $10^7$ electromagnetic traces, measuring the post-quantum Kyber algorithm.

There have been several proposals to use shuffling for neural network implementations.
Nozaki and Yoshikawa~\cite{nozaki2021shuffling}, and Brosch et al.~\cite{brosch2022counteract} proposed a software-based shuffling of multiplications within a neuron to make the parameter recovery infeasible.
The random shuffle algorithm that is commonly used in such cases is the Fisher-Yates algorithm~\cite{fisher1963statistical} as it provides optimal $\mathcal{O}(n)$ time complexity and produces an unbiased permutation.
However, it was shown by Ganesan et al.~\cite{ganesan2023blackjack} that this shuffling method can be easily broken by an SCA targeting the division operation.
They provided a way to reorder the parts of SCA traces back to the original order, thus allowing the parameter recovery.
To counteract this attack, they proposed to realize the shuffling in hardware, avoiding the usage of the Fisher-Yates algorithm.
While their method is secure and efficient, it requires an additional hardware implementation which is not an option for general-purpose microcontrollers.
In this paper, we aim to overcome this limitation by developing a side-channel secure way to realize shuffling in software.

\begin{figure}
    \centering
    \includegraphics[width=0.8\linewidth]{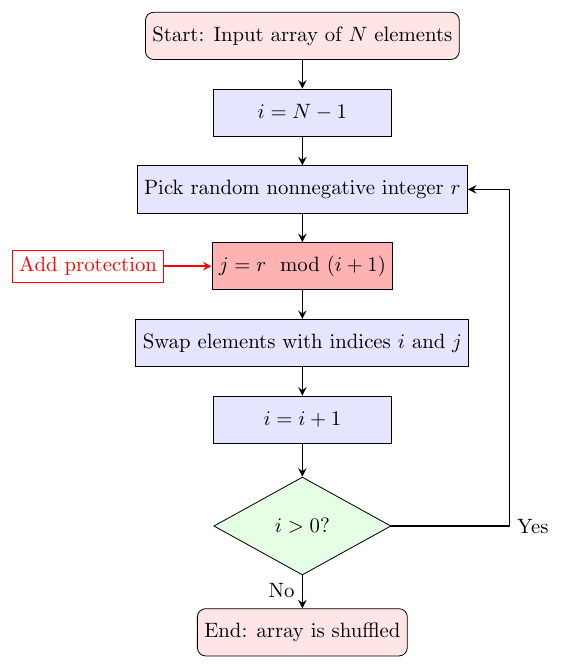}
    \caption{A flow chart depiction of the proposed protected version of the Fisher-Yates algorithm.}
    \label{fig:fisher_yates}
\end{figure}

\vspace{0.2cm}
\textbf{Our contribution.} In this paper, we investigate the possibility of using a shuffling-based countermeasure in software to shuffle the multiplications within a NN layer.
As the original shuffling proposals in~\cite{nozaki2021shuffling} and~\cite{brosch2022counteract} leak the shuffling order via side-channels due to the usage of the Fisher-Yates algorithm~\cite{ganesan2023blackjack}, we alter the algorithm in a way that the attack shown in~\cite{ganesan2023blackjack} is not possible anymore.
Figure~\ref{fig:fisher_yates} shows a high-level depiction of the proposed protection.
The timing overhead over the original method ranges from $4\%$ for a single layer with $100$ neurons and further reduces to $0.49\%$ for a layer with $1000$ neurons.
As the masking itself requires additional storage, the memory overhead comprises two secret arrays of the size of the biggest layer in the network.

To show that the proposed method does not influence the security level of the original Fisher-Yates shuffle, we experimentally evaluate this countermeasure proposal on a small network.
This shows the best-case scenario for the attacker as the number of ways to shuffle such a network is limited.
Our results, performed on an ARM Cortex-M4 processor measured with a ChipWhisperer Lite evaluation platform, show that the proposed shuffling method effectively thwarts the attacker's effort to recover model parameters.

\vspace{0.2cm}
\textbf{Organization.} The rest of this paper is organized as follows.
Section~\ref{sec:background} provides the necessary background and overviews the related work.
Section~\ref{sec:countermeasure} presents the main idea of our shuffling-based countermeasure proposal.
Section~\ref{sec:eval} shows the experimental evaluation results.
Section~\ref{sec:discuss} provides a discussion, and finally, Section~\ref{sec:concl} concludes this paper and provides directions for future work.

\section{Background}
\label{sec:background}
In this section, we will first give an overview of side-channel attacks and countermeasures.
Later, we will detail the current state-of-the-art in the protection of NNs against side-channel attacks.

\subsection{Side-Channel Attacks}
SCAs are a class of cybersecurity threats that exploit indirect information leakage from a device rather than targeting its primary algorithmic or protocol vulnerabilities.
These attacks utilize physical or behavioral characteristics generated during the sensitive computation, such as execution timing, power consumption, electromagnetic emanation, or cache activation to recover secret data.
Originally, SCAs have been proposed for the recovery of cryptographic keys~\cite{kocher1999differential}.
However, they can be used for recovering any kind of secret data used in the computation, such as parameters of machine learning models that are an intellectual property~\cite{batina2019csi}.
Generally, there are two types of SCA analysis methods: \textit{profiled} and \textit{non-profiled}.
In the profiled setting, the attacker is assumed to possess an identical copy of the device under test (DUT), being able to build the model of the device with preliminary measurements.
After this model is created, the attacker can then launch the actual attack on the DUT requiring only a few executions of the sensitive computation in the ideal case.
In the non-profiled setting, the attacker only has the DUT in their possession, and therefore, launches the attack directly without creating the device model.

One of the main non-profiled side-channel analysis methods is correlation power analysis (CPA)~\cite{brier2004correlation}.
In CPA, the attacker first creates a set of hypotheses on a small part of the secret data -- this can be, for example, a byte.
Then, the attacker measures the power consumption traces capturing the part of the execution utilizing that portion of the secret data while varying the non-secret input.
Finally, a statistical correlation is calculated between the hypothetical power consumption (based on the hypotheses of the secret data) and the measured power traces.
If the experimental setting is correct, for the correct hypothesis, the absolute correlation is significantly higher than for incorrect ones.
Apart from being able to recover secret keys of unprotected block cipher implementations, this type of attack has been shown capable of recovering model parameters of neural networks~\cite{batina2022implementation}.
In this paper, we utilize CPA to first show how the attack works on unprotected NN implementation and then to show the effectiveness of the proposed countermeasure.

\subsection{Side-Channel Countermeasures}
There are two broad categories of side-channel countermeasures: \textit{masking} and \textit{hiding}~\cite{hou2024cryptography}.

The goal of masking is to randomize the intermediate values processed by the DUT to make the leakage independent of these values.
For this purpose, we generate a random value, called \textit{mask}, by which the secret intermediate value is concealed using a binary operation.
Masking has been shown to be provably secure under certain conditions~\cite{prouff2013masking}.

The goal of hiding is to either balance or randomize the leakages coming from the DUT to remove the data or operation dependency.
This can be done by various techniques implemented either in software or hardware.
For example, a so-called dual-rail precharge logic utilizes wires carrying complementary values to balance the leakage in hardware.
In software, dummy operations and balanced data encoding~\cite{hou2024another} are some of the approaches.
Generally, hiding-based methods can be overcome with a high number of measurements, however, they can make the attack much harder to execute.
They are also often combined with masking-based countermeasures that suffer from the threat of higher-order CPA attacks~\cite{standaert2005masking}.

\subsection{Related Work}

\subsubsection{SCAs on Neural Networks}
The seminal work by Batina et al.~\cite{batina2019csi} showed how to recover the type of activation function, number of layers and neurons, and weights.
The type of activation function was recovered by a timing attack, the number of layers/neurons by visual inspection of the traces, and the weights by CPA.
The networks were implemented on 8-bit ATmega328P and 32-bit ARM Cortex-M3 microcontrollers.

This was followed by a myriad of works focusing on the practical recovery of various network information implemented on different devices.
To mention a few, \cite{yan2020cache} showed the practical attack on ZYNQ XC7000 SoC on Pynq-Z1 board, \cite{gongye2020reverse} on Intel i7-7700 processor, \cite{chmielewski2021reverse} on Nvidia Jetson Nano, and \cite{kurian2025tpuxtract} on Google Edge TPU.
An embedded OpenVINO framework for implementing models on generic Edge devices was analyzed in~\cite{jap2024side}.
These research efforts show that SCAs on neural network implementations are indeed a practical attack scenario and it is of interest to explore various ways to protect these models.

\subsubsection{Countermeasures against SCAs on Neural Networks}
Masking has been applied to NNs previously in~\cite{dubey2020maskednet}.
They proposed a novel design of components such as masked adder trees and masked ReLUs.
Apart from the need for a pseudo-random number generator to provide the randomness, the design requires twice the latency and needs 2.7$\times$ more look-up tables, 1.7$\times$ more flip-flops, and 1.3$\times$ more BRAM.
The work was later extended in~\cite{dubey2020bomanet} to the entire neural network, resulting in 3.5$\times$ latency overhead and 5.9$\times$ area overhead.
They further improved their masked design in~\cite{dubey2022modulonet} by utilizing modular arithmetic in neural networks, allowing more efficient application of masking.
However, it was later shown that such an implementation can be broken by a heat-induced power leakage~\cite{mehta2024bake}.

Hiding has been utilized in the form of shuffling~\cite{nozaki2021shuffling} and desynchronization~\cite{breier2023desynchronization}.
In~\cite{nozaki2021shuffling} the authors shuffle the order of multiplications within a neuron to drastically lower the success rate of a successful attack.
In~\cite{brosch2022counteract}, the authors shuffle both the neurons within a network and the multiplications within a neuron.
They realize an electromagnetic SCA on an ARM Cortex-M4 microcontroller and provide theoretical estimations on the number of measurements required for a successful attack.
For example, even a small network with three layers of $15, 10, 10$ neurons per layer would require $\approx 47$ million SCA measurements, making the attack impractical.

However, both of the shuffling approaches utilize software shuffling based on the Fisher-Yates algorithm
In~\cite{ganesan2023blackjack} it was shown that the division operation of this algorithm leaks side-channel information that can be used to reorder the shuffled parts of the traces back to their original order.
Instead, they propose a hardware-based shuffling that avoids the use of that algorithm.
Naturally, a hardware solution is efficient and provides a good security level.
However, it requires an additional circuit to perform the shuffle, thus such protection needs to be added during the design phase of the chip and is not applicable to general-purpose hardware.
In this paper, we aim to overcome this limitation by improving the Fisher-Yates algorithm to make it secure against SCA.

\section{Countermeasure Proposal}
\label{sec:countermeasure}
We propose utilizing a protected version of the Fisher-Yates shuffling algorithm to randomize multiplication operations as a countermeasure against CPA attacks.
The standard Fisher-Yates algorithm, widely known for its efficiency in shuffling, is detailed in Algorithm~\ref{alg:Fisher-Yates}.

\begin{algorithm}[tb]
\KwIn{\texttt{array}, $N$\tcp{\texttt{array} contains elements to be shuffled and $N$ is the number of elements in \texttt{array}}}
\KwOut{Shuffled \texttt{array}}
\For{$i=N-1, N-2, \dots, 1$}{
$r=\text{rand}()$\label{line:FY:r}\tcp{rand$()$ generates a random nonnegative integer}
$j=r\mo (i+1)$\label{line:FY:j}\\
swap $\texttt{array}[i]$ and $\texttt{array}[j]$\label{line:FY:swap}\\
}
\Return \texttt{array}
\caption{Fisher-Yates algorithm for shuffling.}
\label{alg:Fisher-Yates}
\end{algorithm}

This algorithm was previously used for shuffling multiplications in neural networks to defend against CPA attacks, as demonstrated in~\cite{brosch2022counteract, nozaki2021shuffling}. 
However, the algorithm’s security was compromised in~\cite{ganesan2023blackjack}.
In this paper, the authors demonstrated that when a division operation is performed, the values of the dividend and divisor can be deduced by analyzing power variations.
Since division is employed during the computation of modular reduction in line~\ref{line:FY:j} of the algorithm, the value of $j$ can be recovered. 
Consequently, each swapping operation in line~\ref{line:FY:swap} can be reversed, allowing the attacker to reorder segments of traces corresponding to individual multiplication operations and align them correctly.

To mitigate these vulnerabilities, we propose a protected Fisher-Yates algorithm that integrates masking techniques for modular reduction and Blakely's method for modular multiplication~\cite{blakely1983computer}. 
The masking conceals the value of $j$, while Blakely's method eliminates the need for division in modular reduction, thereby enhancing the algorithm's resistance to the attack presented in~\cite{ganesan2023blackjack}.

Next, we provide a detailed explanation of Blakely's modular multiplication method in Subsection~\ref{sec:blakely}, followed by an analysis of how it can be combined with a multiplicative masking technique to compute $j$ in line~\ref{line:FY:j}, as discussed in Subsection~\ref{sec:masking}.
Finally, we present the design of our proposed countermeasure tailored for neural network computations in Subsection~\ref{sec:shuffling}.

\subsection{Blakely's method for modular multiplication}
\label{sec:blakely}
Let $n\geq2$ be an integer.
Consider two integers $a,b$ such that $0\leq a,b<n$.
The binary representation of $a$ can be expressed as
\[
a=a_{\ell_a-1}a_{\ell_a-2}\cdots a_1a_0,
\]
where $\ell_a$ is the bit length of $a$.
The product $ab$ can then be computed as follows
\begin{equation}\label{eq:pro}
    ab=\left(\sum_{i=0}^{\ell_a-1} a_i2^i\right)b=\sum_{i=0}^{\ell_a-1}2^ia_ib.
\end{equation}
Blakely's method for computing the modular multiplication $ab\mo n$ leverages this representation. 
The step-by-step details are provided in Algorithm~\ref{alg:blakley}.

\begin{algorithm}[htb]
\KwIn{$n,\ a,\ b$\tcp{$n\in\ZZ, n\geq2$; $0\leq a,b<n$}}
\KwOut{$ab\mo n$}
$\ell_a=$ bit length of $a$\\
$R=0$\\
\For{$i=\ell_a-1$, $i\geq0$, $i--$
 \label{line:alg:blakelybit1:loop}}
 	{
 	$R=2R+a_ib$\label{line:alg:blakelybit1:aib}\tcp{compute $ab$}
        \If{$R\geq n$\label{line:alg:blakely:if}}{$R=R-n$}
        \If{$R\geq n$}{$R=R-n$\label{line:alg:blakely:modn}}
  	}
  	\Return $R$
	\caption{\texttt{Blakely}, Blakely's method for computing modular multiplication.\label{alg:blakley}}
\end{algorithm}

In the Algorithm, line~\ref{line:alg:blakelybit1:aib} computes the product $ab$ as described in Equation~(\ref{eq:pro}).
Subsequently, lines~\ref{line:alg:blakely:if} --~\ref{line:alg:blakely:modn} calculate $R\mo n$.
Specifically, when $i=\ell_a-1$, in line~\ref{line:alg:blakelybit1:loop}, we have
\[
R=a_ib\leq n-1.
\]
In the following iterations of the loop at line~\ref{line:alg:blakelybit1:loop}, the value of $R$ becomes
\[
R=2R+a_ib\leq 2(n-1)+(n-1)=3n-3.
\]
Thus, by comparing $R$ with $n$ twice, $R\mo n$ can be efficiently computed. 
This approach avoids performing division operations and is therefore resistant to the attack described in~\cite{ganesan2023blackjack}.

\subsection{Masked Shuffling}
\label{sec:masking}
To make the shuffling algorithm secure, we introduce masking for the modular reduction operation in line~\ref{line:FY:j} of Algorithm~\ref{alg:Fisher-Yates} utilizing two secret arrays.
The arrays, $S_1$ and $S_2$, are generated such that $S_1$ contains random positive integers $S_1[k]$ coprime with $k+3$, while $S_2[k]$ represents the corresponding multiplicative inverse of $S_1[k]$ modulo $k+3$.
More specifically, for $k=0,1,\dots,N-3$, the arrays satisfy the following properties:
\begin{equation}\label{eq:s1}
    \gcd(S_1[k],k+3)=1,\quad S_1[k] >0,
\end{equation}
and
\begin{equation}\label{eq:s2}
 S_2[k] = S_1[k]^{-1}\mo (k+3),
\end{equation}
where $N$ denotes the total number of elements to be shuffled.
In particular, the following condition holds:
\[
1\leq S_2[k]\leq k+2,\quad k=0,1,\dots,N-3.
\]

The details of our protected Fisher-Yates shuffling algorithm are outlined in Algorithm~\ref{alg:protected}.  
In this algorithm, $r$ from line~\ref{line:protected:r} and $r'$ from line~\ref{line:protected:r1} are both random nonnegative integers.  
Line~\ref{line:protected:t} computes 
\[
t = (rS_1[i-2]+r'(i+1))\mo (i+1)=rS_1[i-2] \mo(i+1),
\]
where by design, $S_1[i-2]$ is a random integer coprime with $i+1$ (see Equation~(\ref{eq:s1})).
From Equation~(\ref{eq:s2}), $S_2[i-2]$ is the multiplicative inverse of $S_1[i-2]$ modulo $(i+1)$, we have
\[
S_1[i-2]S_2[i-2]\mo (i+1)= 1.
\]
Thus, using Blakely's method, line~\ref{line:protected:j} computes
\begin{eqnarray*}
    j &=& tS_2[i-2]\mo (i+1)\\
    &=&rS_1[i-2]S_2[i-2]\mo (i+1)=r\mo(i+1)
\end{eqnarray*}
We have shown that lines~\ref{line:protected:t} and~\ref{line:protected:j} in Algorithm~\ref{alg:protected} implement
\[
j=r\mo (i+1).
\]
Consequently, lines~\ref{line:protected:r} to~\ref{line:protected:swap} in Algorithm~\ref{alg:protected} correspond to lines~\ref{line:FY:r} to~\ref{line:FY:swap} in Algorithm~\ref{alg:Fisher-Yates} for $i=N-1, N-2,\dots, 2$.
It is easy to see that lines~\ref{line:protected:ri1} to~\ref{line:protected:swapi1} in Algorithm~\ref{alg:protected} implement lines~\ref{line:FY:r} to~\ref{line:FY:swap} from Algorithm~\ref{alg:Fisher-Yates} for the case $i=1$.

\subsection{Security of the implementation against the attack in~\cite{ganesan2023blackjack}}
We have established that Blakely's method for computing modular multiplication is secure against the attack presented in~\cite{ganesan2023blackjack}, as it does not involve any division in its computation.  
In the context of Algorithm~\ref{alg:protected}, the only potentially vulnerable line is line~\ref{line:protected:t}.  
According to the attack described by the authors, the attacker can recover the values of $t$ and
\[
rS_1[i-2]+r'(i+1).
\]

However, knowing the value of $t$ does not help the attacker in recovering the value of $j$, since $j$ is computed by multiplying $t$ with a random number modulo $i+1$.
Furthermore, as $S_1[i-2]$ is a secret value and both $r$ and $r'$ are random numbers, knowledge of $rS_1[i-2]+r'(i+1)$ also provides no useful information to the attacker. 

The shuffling algorithm is executed prior to each inference computation.
Consequently, each time, the attacker can recover few other values $r'S_1[i-2]+r'_1(i+1)$.
In general, the attacker could obtain a series of values:  
\[
\lambda_1S_1[i-2]+\beta_1,\ \lambda_2S_1[i-2]+\beta_2,\ \dots
\]
where the $\lambda$s and $\beta$s are random numbers.
It is worth noting that without the $\beta$s, the attacker might be able to deduce $S_1[i-2]$ by computing the greatest common divisor (GCD) of the values
\[
\lambda_1S_1[i-2],\ \lambda_2S_1[i-2],\ \dots
\]
However, with the $\beta$s being added to each value, the recovered values no longer provide any meaningful information about $S_1[i-2]$.
Consequently, no information about $S_2[i-2]$ is revealed, which is essential for determining the value of $j$. 

\begin{algorithm}[htb]
\KwIn{\texttt{array}, $N$, $S_1$, $S_2$\tcp{\texttt{array} contains elements to be shuffled and $N$ is the number of elements in \texttt{array}; $S_1$ and $S_2$ are given by Equations~(\ref{eq:s1}) and~(\ref{eq:s2}) respectively}}
\KwOut{Shuffled \texttt{array}}
\For{$i=N-1, N-2, \dots, 2$}{
$r=\text{rand}()$\label{line:protected:r}\tcp{rand$()$ generates a random nonnegative integer}
$r'=\text{rand}()$\label{line:protected:r1}\\
$t = (rS_1[i-2]+r'(i+1))\mo (i+1)$\label{line:protected:t}\\
$j=\texttt{Blakely}(i+1,t,S_2[i-2])$\label{line:protected:j}\tcp{Compute $tS_2[i-2]\mo (i+1)$ using Blakely's method in Algorithm~\ref{alg:blakley}}
swap $\texttt{array}[i]$ and $\texttt{array}[j]$\label{line:protected:swap}\\
}
$r=\text{rand}()$\label{line:protected:ri1}\\
$j=r\mo 2$\\
swap $\texttt{array}[1]$ and $\texttt{array}[j]$\label{line:protected:swapi1}\\
\Return \texttt{array}
\caption{Protected Fisher-Yates algorithm for shuffling.}
\label{alg:protected}
\end{algorithm}

\subsection{Shuffling multiplication}
\label{sec:shuffling}
Before each inference computation, we generate a shuffled array for each layer of the network as follows. 
Let \( N \) represent the number of input neurons in the layer, and define:  
\[
\texttt{array} = \{1, 2, \dots, N\},
\]  
where each element corresponds to a specific input neuron. 
The \texttt{array} is shuffled using Algorithm~\ref{alg:protected}, and the resulting shuffled sequence of indices is used to reorder the multiplication operations in the layer's computation during inference. 
Specifically, for each output neuron in the layer, the multiplications are shuffled independently using the same shuffled \texttt{array}.

It is worth noting that it is unnecessary to create separate secret arrays \( S_1 \) and \( S_2 \) for each layer. 
Instead, \( S_1 \) and \( S_2 \) can be computed based on the maximum number of neurons in any layer of the network. 
The shuffling for other layers can then utilize the relevant entries from \( S_1 \) and \( S_2 \) to perform the protected shuffling.

To evaluate the security of the implementation against the attack proposed in~\cite{ganesan2023blackjack}, one might argue that brute-forcing the values in $S_2$ is feasible, given that $S_2[k]$ is known to be between $1$ and $k+2$.
More specifically, since $S_2[k]$ is coprime with $k+3$, the total number of possible values for $S_2[k]$ is $\varphi(k+3)$, where $\varphi$ denotes Euler's totient function. For an integer $n \geq 2$,
\[
\varphi(n) = \left\vert\{a\ |\ a\in\ZZ,\ 1\leq a\leq n-1,\  \gcd(a,n)=1\}\right\vert.
\]
Consequently, the total number of possible values in $S_2$ is given by:
\[
\prod_{k=0}^{N-3}\varphi(k+3)=\prod_{k=3}^{N}\varphi(k).
\]
For example, when $N=20$, the number of possible values in $S_2$ is approximately
\[
\prod_{k=3}^{20}\varphi(k)\approx2^{45}.
\]

To execute a brute-force attack, the attacker must recover all the weight values to test whether the output for a given input matches (or approximates) a known correct output of the network. 
This involves re-shuffling the multiplications for each neuron in the hidden layer for every trace. 
The attacker would then use the recovered weights to deduce the next layer’s weights iteratively, continuing this process until the output layer’s weights are fully recovered. 
Assuming each attack requires at least $1$ second, the time needed to brute-force all the possible values in $S_2$ for a network with at most $20$ neurons in each layer would be approximately:
\[
2^{45}\text{ seconds }\approx 2^{20} \text{ years}.
\]
This duration is clearly impractical, particularly considering the rapid advancements in AI. 
It is reasonable to anticipate that within a few years (or months), the secret AI model attacker is trying to recover will likely be replaced by a more advanced and efficient one, rendering such an attack even less relevant.

\section{Experimental Evaluation}
\label{sec:eval}
As in~\cite{batina2019csi}, we adopt an approach to recover the different components of a secret weight individually.  
We begin by recalling the $32-$bit (single-precision) floating-point representation as defined by the IEEE 754 standard. Specifically, the binary string $b_{31}b_{30}\cdots b_0$ represents the floating-point number: 
\[
(-1)^{b_{31}}\times 2^{b_{30}b_{29}\cdots b_{23}-127}\times1.b_{22}b_{21}\dots b_0,
\]
where $b_{30}b_{29}\cdots b_{23}$ denotes the integer
\[
b_{30}2^{8}+b_{29}2^{7}+\dots +b_{23}
\]
and $1.b_{22}b_{21}\cdots b_0$ represents the value
\[
1+\frac{b_{22}}{2}+\frac{b_{21}}{2^{2}}\dots+\frac{b_0}{2^{23}}.
\]
In this representation, \( b_{31} \) is referred to as the \textit{sign bit}, \( b_{30}b_{29}\cdots b_{23} \) as the \textit{exponent}, and \( b_{22}b_{21}\cdots b_0 \) as the \textit{mantissa}. 
Additionally, we refer to \(b_{22}b_{21}\cdots b_{16}\), \(b_{15}b_{14}\cdots b_{7}\), and \(b_{6}b_{5}\cdots b_{0}\) as the first byte, second byte, last seven bits of the mantissa.

Although similar approaches have been used in various works for recovering secret weight values during neural network computations, the specific CPA attack steps and successful recovery of the sign bit and exponent bits have not been comprehensively detailed in the literature.  

\begin{algorithm}[tb]
\KwIn{$r$, $W$, $M_w$, $q$\tcp{$r$ is the array of absolute correlations obtained in Step~\ref{step:correlations}; 
$W$ is the set of all possible weight values from Step~\ref{step:weight};
$M_w$ is the total number of weights in $W$;
$q$ is the number of time samples in the target range identified in Step~\ref{step:range}}}
\KwOut{Absolute value of correlation coefficients for each exponent value}
\textbf{array} of size $256\times q$ $r_e$ \\
Initialized all entries of $r_e$ to $0$\\
\For{$j=0,1,\dots,M_w-1$}{
Let $b^j_{31}b^j_{30}\cdots b^j_0$ be the binary representation of $w_j$\\
$\texttt{exponent} = b^j_{30}2^{8}+b^j_{29}2^{7}+\dots +b^j_{23}$\\
\For{$t=0,1,\dots,q-1$}{\label{line:recovery:t}
\If{$r_e[\texttt{exponent}]<r[j,t]$\label{line:recovery:if}}{
$r_e[\texttt{exponent}]=r[j,t]$\label{line:recovery:re}
}
}
}
  	\Return $r_e$
	\caption{Compute absolute correlations for each possible exponent value.\label{alg:recovery}}
\end{algorithm}

To ensure the completeness of our evaluation, in Subsection~\ref{sec:steps}, we will outline the CPA attack methodology for recovering the different components of a secret weight value, ultimately reconstructing the entire weight.  
Subsequently, in Subsection~\ref{sec:results}, we will demonstrate how the different components of a secret weight value can be successfully recovered in an unprotected implementation, as well as how our proposed countermeasure effectively mitigates such an attack.  

\subsection{CPA Attack Steps}
\label{sec:steps}
Below, we detail the steps to recover the first secret weight value of the first neuron in the first hidden layer.
Other weight values can be recovered in a similar manner.
\begin{enumerate}[label=\Circled{\arabic*}]
\item \label{step:traces}\textbf{Collect attack traces.}
    To target the first secret weight value, traces are collected with all inputs fixed except for the first neuron in the input layer.  
    Let \(\mathcal{T} = \{l_0, l_1, \dots, l_{M_L-1}\}\) represent the set of \(M_L\) attack traces.  
    For each trace \(l_i\), \(a_i\) (\(i = 0, 1, \dots, M_L-1\)) denotes the corresponding random input to the first neuron. 
\item\label{step:range} \textbf{Identify target range of time samples.}  
    As demonstrated before, visual inspection of traces can reveal the time samples corresponding to the first multiplication computation for unprotected implementations~\cite{batina2019csi}.  
    In the case of shuffled implementations, the precise location of the multiplication is unknown, so the entire multiplication computation segment for the first hidden neuron is considered as the target duration.  
    Let \(q_s\) and \(q_e\) denote the start and end time samples of the identified range, with \(q := q_e - q_s + 1\) representing the total number of time samples in this range. 
\item \label{step:weight} \textbf{Compute hypothetical weights.}  
    Given the vast number of possible weight values, it is impractical to consider every single value.  
    To overcome this issue, we first determine the desired precision, defined by the number of decimal places to be recovered.
    A reasonable range for the weight values can also be assumed, same as in~\cite{batina2019csi}.  
    Let \(W = \{w_0, w_1, \dots, w_{M_w-1}\}\) represent the set of all possible weight values. 
\item \textbf{Compute hypothetical leakages.}  
    For each weight value \(w_j\), the \textit{hypothetical} leakage for input \(a_i\) is computed as the Hamming weight (HW) of the product \(w_j a_i\) -- the number of $1$s in the binary representation of the $32-$bit floating-point number \(w_j a_i\), as defined by the IEEE 754 standard.  
    Specifically, construct the matrix \(\mathcal{H}\) of size \(M_w \times M_L\) such that  
    \[
    \mathcal{H}[j,i] = \text{HW}(w_j a_i),
    \]  
    for \(j = 0, 1, \dots, M_w-1\) and \(i = 0, 1, \dots, M_L-1\).  
\item \label{step:correlations} \textbf{Compute correlations.}  
    For each weight value \(w_j\) and each time sample \(t\) in the target range, calculate the absolute value of the correlation coefficient between the hypothetical and real leakages.  
    Let \(\mathcal{L}\) denote the array of real leakages in the target range:  
    \[
    \mathcal{L}[t,i] = l_i[t + q_s],
    \]  
    for \(i = 0, 1, \dots, M_L-1\) and \(t = 0, 1, \dots, q-1\).  
    Compute the \(M_w \times q\) matrix \(r\) such that  
    \[
    r[j, t] = \left\vert\frac{\sum_{i=0}^{M_L-1} (\mathcal{H}[j,i] - \overline{\mathcal{H}[j]})(\mathcal{L}[t,i] - \overline{\mathcal{L}[t]})}{\sqrt{\var{\mathcal{H}[j]} \var{\mathcal{L}[t]}}}\right\vert,
    \]  
    where \(\overline{\mathcal{H}[j]}\) (resp. \(\overline{\mathcal{L}[t]}\)) and \(\var{\mathcal{H}[j]}\) (resp. \(\var{\mathcal{L}[t]}\)) represent the mean and variance of the values in the $j$th (resp. $t$th) row of $\mathcal{H}$ (resp. $\mathcal{L}$)
\item \textbf{Recovery of weight value.}
    As previously mentioned, the recovery process involves extracting different parts of the weight separately. 
    For example, the algorithm for recovering the exponent bits of the weight is detailed in Algorithm~\ref{alg:recovery}.
    For each time sample (line~\ref{line:recovery:t}), we record the highest absolute value of the correlation coefficient for each exponent value (lines~\ref{line:recovery:if} and~\ref{line:recovery:re}) in the array $r_e$.
    To determine the exponent bits, we plot the absolute correlations corresponding to each exponent value against all $q$ time samples.  
    The exponent value achieving the highest peaks is identified as the correct one.  
    Similarly, the sign bit, the first byte, second byte, and the last seven bits of the mantissa can be recovered using the same approach. 
\end{enumerate}

\subsection{Experimental Setup}
\label{sec:setup}
We used ChipWhisperer Lite with ARM Cortex-M4 (STM32F3) target as the basis for our setup.
The target frequency was set to $7.3728$ MHz, and the ADC frequency was set to $4\times$ that.
The value shown in the y-axis of the power traces is proportional to the current going through the shunt resistor --- we refer to this as ``leakage'' in the graphs as these are not exact current values.

\subsection{Results}
\label{sec:results}


\begin{figure}[tb]
\centering
\captionsetup[subfloat]{labelfont=small,textfont=small}
\subfloat[\textnormal{Unprotected network.}\label{fig:onelayer_unprotected}] {\includegraphics[width=0.48\textwidth]{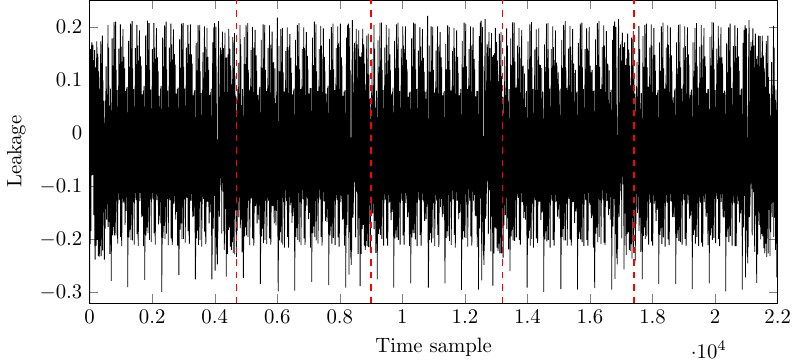}}\vfill
\subfloat[\textnormal{Protected network.}\label{fig:onelayer_protected}]{\includegraphics[width=0.48\textwidth]{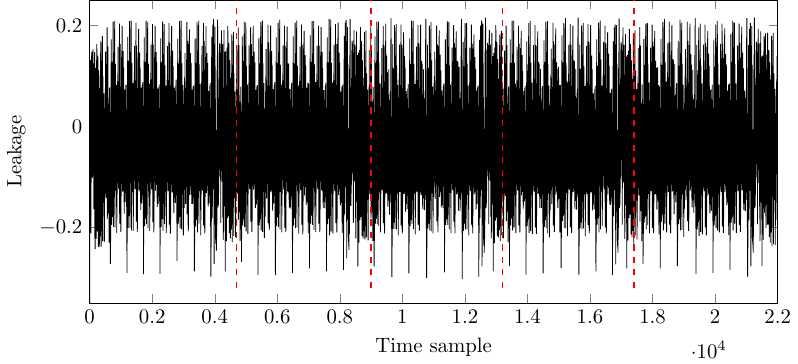}}
\caption{Power traces corresponding to the computation of the first hidden layer in (a) unprotected and (b) protected implementations. 
The durations of each neuron computations are clearly distinguishable in both cases as indicated by red dotted lines.} \label{fig:onelayer}
\end{figure}

\begin{figure}[tb]
\centering
\captionsetup[subfloat]{labelfont=small,textfont=small}
\subfloat[\textnormal{Unprotected network.}\label{fig:oneneuron_unprotected}] {\includegraphics[width=0.48\textwidth]{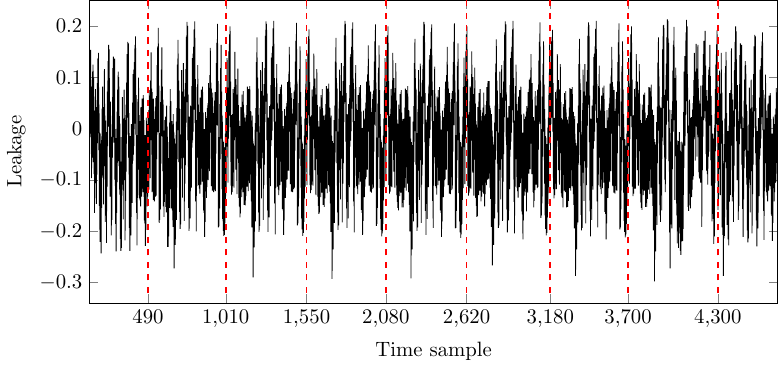}}\vfill
\subfloat[\textnormal{Protected network.}\label{fig:oneneuron_protected}]{\includegraphics[width=0.48\textwidth]{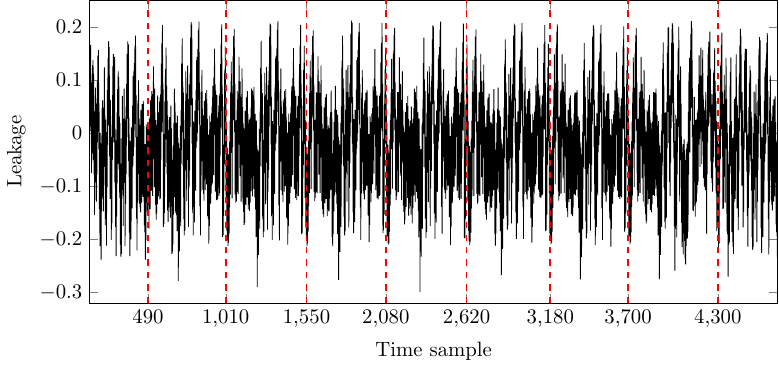}}
\caption{Power traces for the first neuron computation in the first hidden layer for (a) unprotected and (b) protected implementations. In both cases, the durations of multiplication operations are distinguishable (red dotted lines). 
However, in the unprotected network, the first multiplication (time samples $490$–$1010$) corresponds to the first input neuron, while in the protected case, this correspondence is obscured.} \label{fig:oneneuron}
\end{figure}

To demonstrate our approach, we evaluate it using a small multilayer perceptron (MLP) comprising four layers with $7, 5, 4, 3$ neurons, respectively.
The hidden layers utilize ReLU activation functions, while the output layer employs a sigmoid activation function.
The weight values are randomly generated within the range $[-2, 2]$ with a precision of up to two decimal places.

The power traces for the first hidden layer computations are illustrated in Figure~\ref{fig:onelayer} for both the unprotected network and the network with shuffled multiplications. As shown, the computations for the five neurons are clearly distinguishable in both cases, marked by the red dotted lines.
The visual inspection allowing to determine the number of layers and neurons per layer has been described in~\cite{batina2019csi}.
As the patterns for multiplications and activation functions are clearly distinguishable from each other, this allows the attacker to split the power trace into segments, each containing one layer.
Within the layer trace segment, the attacker can then zoom in to identify the repeating multiplication patterns.

These observations indicate that recovering the network architecture (e.g., the number of neurons per layer) through visual inspection of the power traces remains feasible in both scenarios. 
This is expected as our countermeasure is not designed to conceal the network architecture but rather to protect the secret weight parameters from being extracted using CPA.
We would like to note that this allows a stronger attacker model.

To recover the first secret weight, we zoom into the computation of the first neuron. 
The resulting plots are depicted in Figure~\ref{fig:oneneuron}.
For both unprotected and protected networks, the seven multiplications are distinguishable, as marked by the red dotted lines. 
However, in the unprotected network, the first multiplication (occurring between time samples $490$ and $1010$) corresponds to the first input neuron, whereas in the protected case, the correspondence between multiplications and input neurons are unknown.

\begin{figure*}[tb]
    \centering \small
    \begin{tabular}{ccc}
        \includegraphics[width=0.3\linewidth]{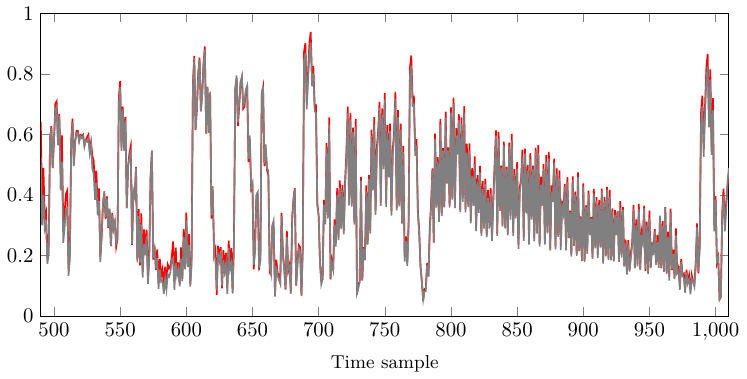} & \includegraphics[width=0.3\linewidth]{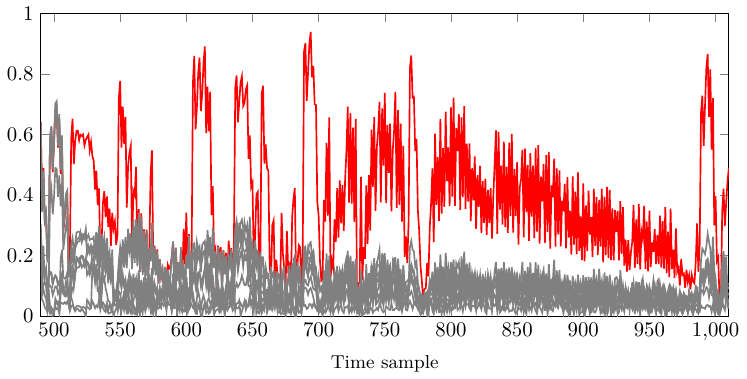} & \includegraphics[width=0.3\linewidth]{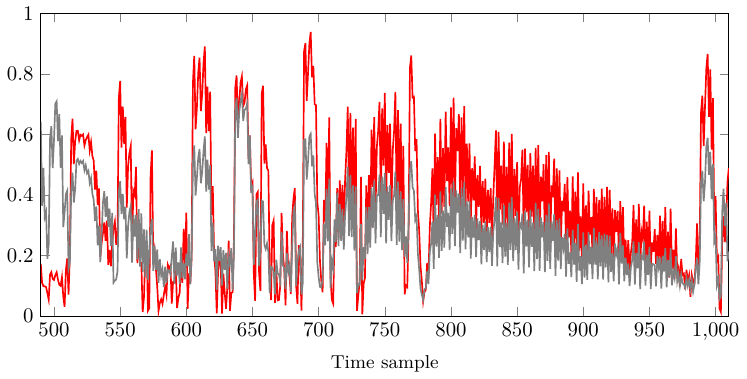} \\
        (a) sign bit & (b) exponent bits & (c) first byte of mantissa
    \end{tabular}
    \begin{tabular}{cc}
    \includegraphics[width=0.3\linewidth]{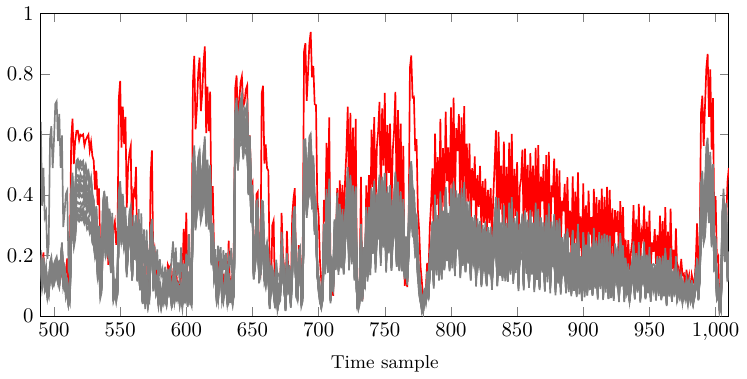} & \includegraphics[width=0.3\linewidth]{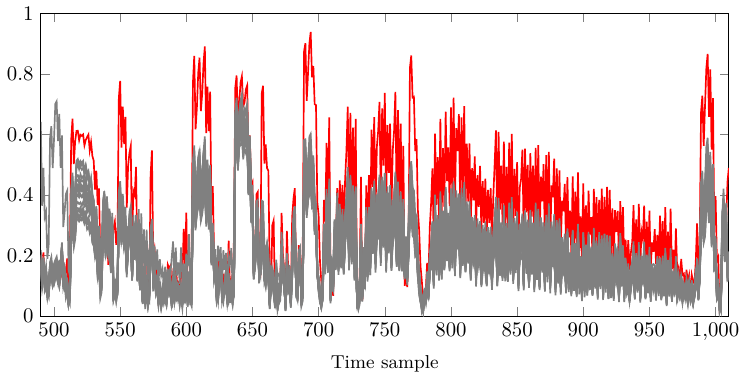} \\
       (d) second byte of mantissa & (e) last seven bits of mantissa
    \end{tabular}
    \begin{tabular}{c}
    \includegraphics[width=0.5\linewidth]{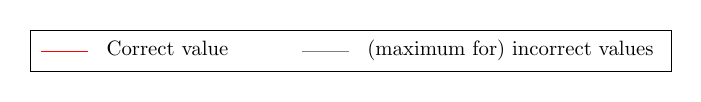}
    \end{tabular}
    \vspace*{-0.3cm}
    \caption{CPA attack results for the unprotected implementation. 
    The $y$-axis represents the absolute correlation. 
    The red lines correspond to the correct values associated with the correct weight of $1.43$, while the gray lines correspond to incorrect values.}
    \label{fig:cpa_unprotected}
\end{figure*}

\begin{figure*}[tb]
    \centering \small
    \begin{tabular}{ccc}
        \includegraphics[width=0.3\linewidth]{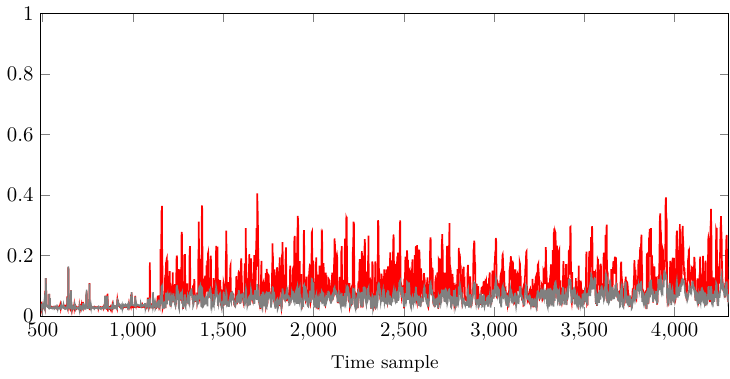} & 
        \includegraphics[width=0.3\linewidth]{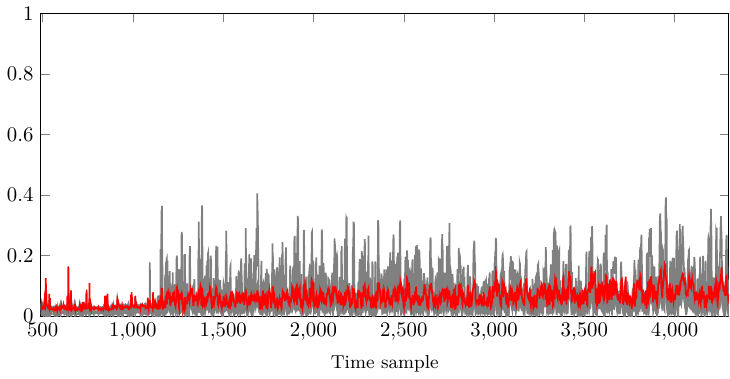} & 
        \includegraphics[width=0.3\linewidth]{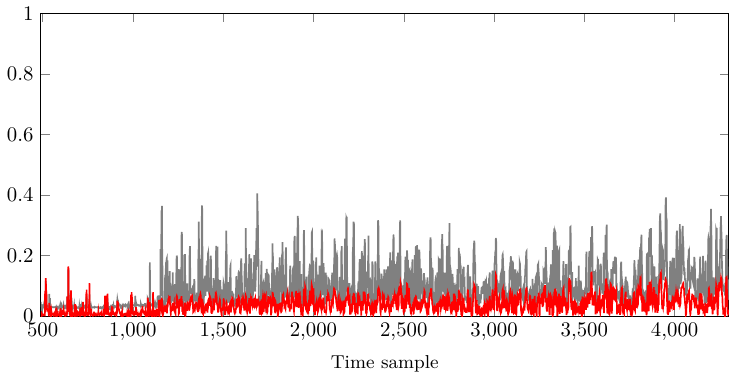} \\
        (a) sign bit & (b) exponent bits & (c) first byte of mantissa
    \end{tabular}
    \begin{tabular}{cc}
    \includegraphics[width=0.3\linewidth]{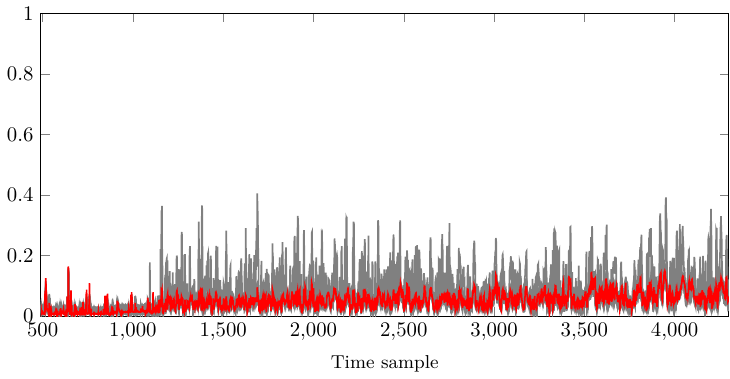} & \includegraphics[width=0.3\linewidth]{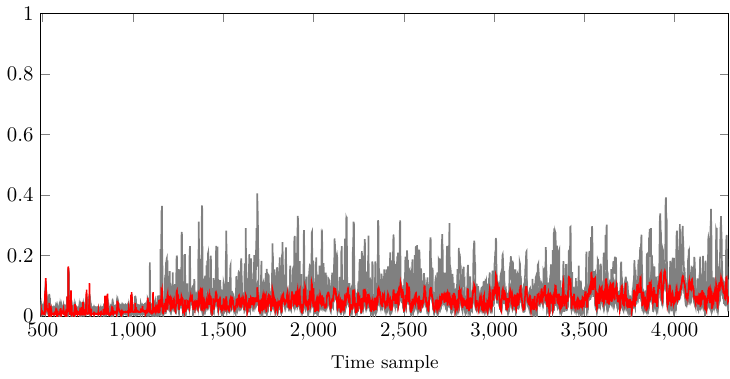} \\
       (d) second byte of mantissa & (e) last seven bits of mantissa
    \end{tabular}
    \begin{tabular}{c}
    \includegraphics[width=0.5\linewidth]{fig/legend.pdf}
    \end{tabular}
    \vspace*{-0.3cm}
    \caption{CPA attack results for the protected implementation. 
    The $y$-axis represents the absolute correlation. 
    The red lines correspond to the correct values associated with the correct weight of $1.43$, while the gray lines correspond to incorrect values.}
    \label{fig:cpa_protected}
\end{figure*}

To perform CPA attacks, we follow the steps outlined in Subsection~\ref{sec:steps}. 
For trace collection in Step~\ref{step:traces}, the first neuron inputs are randomly generated within $[-2, 2]$, while all other input values are set to $0.5$. 
This value was chosen arbitrarily to ensure a nonzero value\footnote{As suggested in~\cite{brosch2022counteract}, the authors introduced a countermeasure to prevent other inputs from being fixed at \(0\), as this would otherwise significantly simplify the attack.}.
We collect $M_L=2000$ traces for the unprotected implementation and $M_L=10,000$ for the protected implementation. 
The weight values are randomly generated within $[-2, 2]$ with a precision of $0.01$, resulting in the set:
\[
    W = \{-2, -1.99, -1.98, \dots, -0.01, 0, 0.01, \dots, 1.99, 2\}.
\]
Following the notation from Step~\ref{step:weight}, the number of hypothetical weight values is $M_w = 401$.

For Step~\ref{step:range}, the target time sample range is identified as follows: for the unprotected implementation, as observed in Figure~\ref{fig:oneneuron}(a), the first multiplication occurs between time samples $q_s=490$ and $q_e=1010$. 
For the protected implementation, the start time sample remains $q_s=490$, but due to shuffling, the range extends to $q_e=4300$ (see Figure~\ref{fig:oneneuron}(b)), which corresponds to the end of the seventh multiplication.

The weight value we used for the first input neuron is $1.43$, corresponding to a sign bit of $0$, an exponent value of $127$, and mantissa bytes as follows: the first byte is $110$, the second byte is $20$, and the last seven bits represent the value $61$. 

The CPA attack results for the unprotected and protected implementations are presented in Figures~\ref{fig:cpa_unprotected} and~\ref{fig:cpa_protected} respectively, where the absolute correlations for the correct values are highlighted in red, while those for incorrect values are shown in gray. 
For the first byte of the mantissa, only the maximum absolute correlation among incorrect values for each time sample is plotted due to the large number of possible values\footnote{Since the weight values are constrained to the range \([-2,2]\) with two decimal places of precision, not all possible values for the exponent and mantissa bytes can be represented, leading to variations in the number of incorrect hypotheses plotted.}.

For the attacks on the unprotected implementation, Figure~\ref{fig:cpa_unprotected} clearly demonstrates that the correct weight value can be successfully recovered.
As detailed in Algorithm~\ref{alg:recovery}, the absolute correlation values were grouped differently depending on the specific weight component being targeted.
For each time sample, the highest absolute correlation within each group was selected and plotted.
Notably, all red lines in the figures exhibit the same behavior, indicating that the true weight value consistently achieves the highest absolute correlation, regardless of how the correlation coefficients are grouped.
Furthermore, the minimal differences between the red and gray lines in Figure~\ref{fig:cpa_unprotected} (a) highlight the importance of recovering each weight component separately. 
In fact, if we plot the absolute correlation for the true weight alongside the highest absolute correlation among incorrect weight hypotheses, the resulting plot closely resembles Figure~\ref{fig:cpa_unprotected} (a), making it challenging to distinguish the true weight from the incorrect ones.

For attacks on the protected implementation, Figure~\ref{fig:cpa_protected}(a) shows that the sign bit of the weight can still be recovered. 
Unlike Figure~\ref{fig:cpa_unprotected}(a), where the exact location of the target multiplication is known for each execution, in this case, the precise timing is unknown. As a result, all seven multiplications were analyzed for each execution.  
Among these, only one multiplication varies per execution, while the remaining six involve three fixed positive weights, three fixed negative weights, and six fixed positive inputs (0.5). Since the majority of these values are positive and appear randomly within the analyzed interval, we believe this explains why the positive sign exhibits a higher absolute correlation with the leakage.
We would like to note that recovering the sign bit itself does not significantly reduce the search complexity of the entire floating point value.
For other components of the weight value, the peaks corresponding to the correct values are significantly lower compared to those of incorrect values.

Moreover, in all cases, the absolute correlations for the protected implementation remain far from $1$, in contrast to the results for the unprotected implementation. 
These findings demonstrate that the CPA attack is ineffective in recovering the target weight value in the presence of our proposed countermeasure.

\subsection{Impact of Increasing the Number of Traces on Attack Performance}
To assess how the attack evolves as the number of traces increases, we conducted the same attack using varying numbers of traces for both protected and unprotected implementations.

\begin{figure*}[tb]
    \centering \small
    \begin{tabular}{ccc}
        \includegraphics[width=0.3\linewidth]{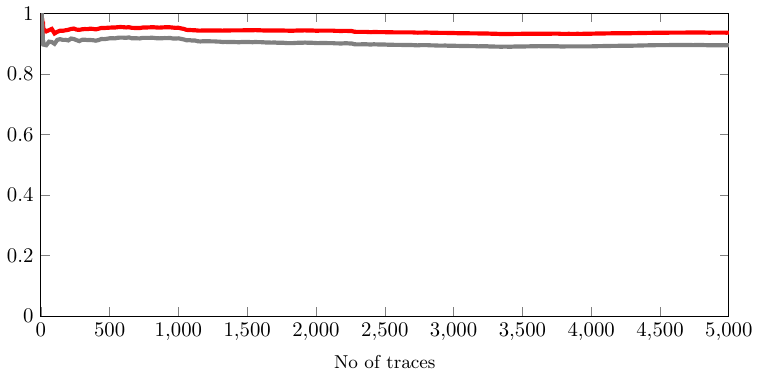} & \includegraphics[width=0.3\linewidth]{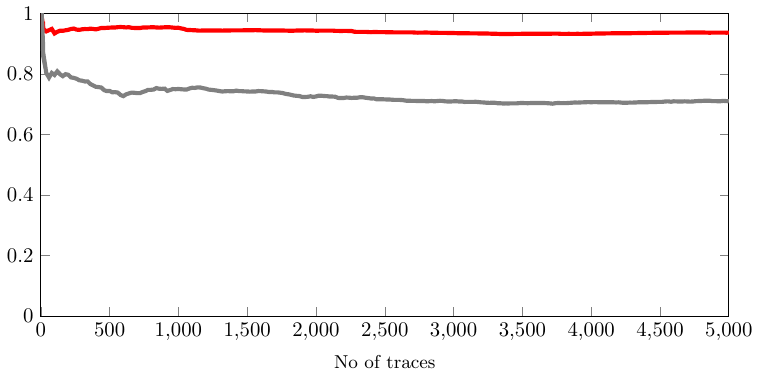} & \includegraphics[width=0.3\linewidth]{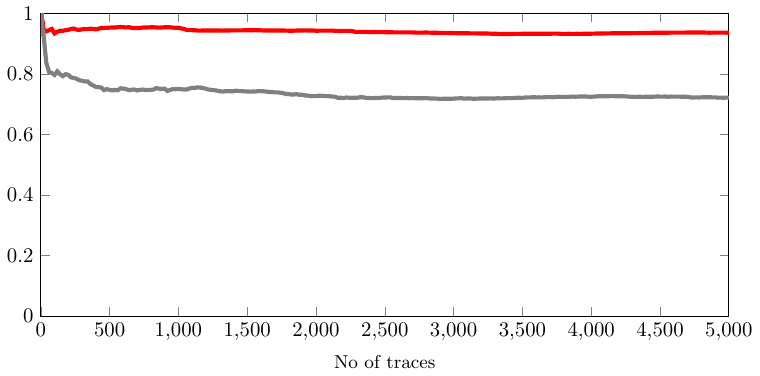} \\
        (a) sign bit & (b) exponent bits & (c) first byte of mantissa
    \end{tabular}
    \begin{tabular}{cc}
    \includegraphics[width=0.3\linewidth]{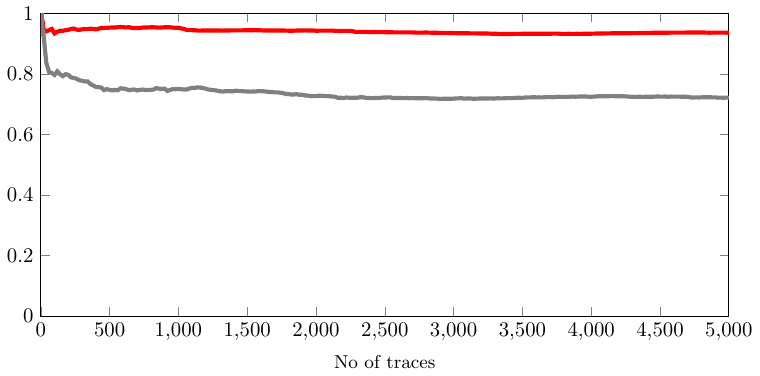} & \includegraphics[width=0.3\linewidth]{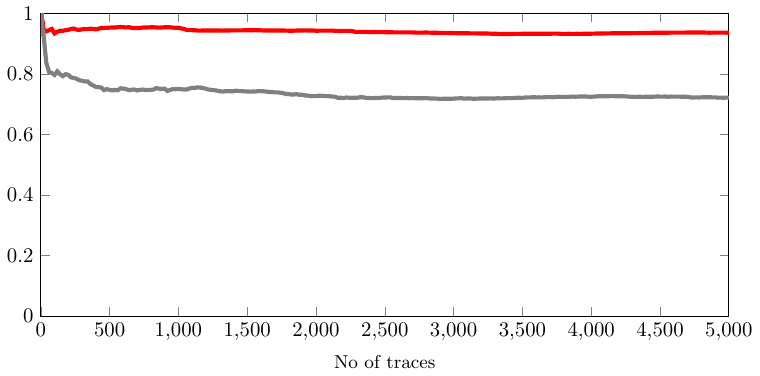} \\
       (d) second byte of mantissa & (e) last seven bits of mantissa
    \end{tabular}
    \begin{tabular}{c}
    \includegraphics[width=0.4\linewidth]{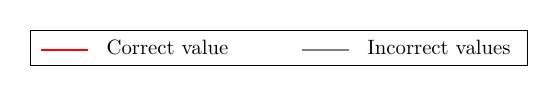}
    \end{tabular}
    \vspace*{-0.3cm}
    \caption{CPA attack results with different numbers of traces for the unprotected implementation. The $y$-axis represents the absolute correlation. The red lines correspond to the correct values associated with the correct weight of $1.43$, while the gray lines correspond to incorrect values.}
    \label{fig:cpa_unprotected_traces}
\end{figure*}

\begin{figure*}[tb]
    \centering \small
    \begin{tabular}{ccc}
        \includegraphics[width=0.3\linewidth]{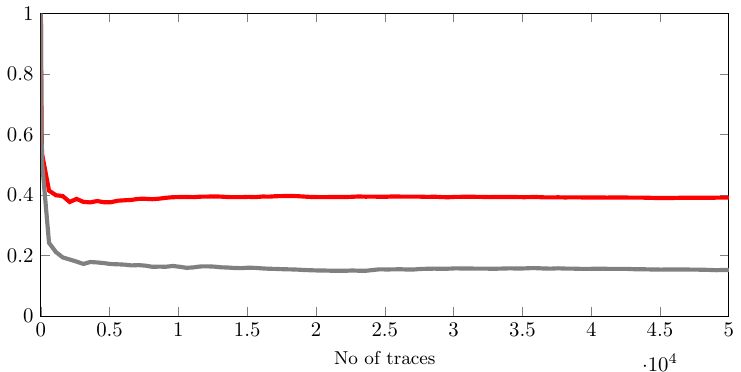} & \includegraphics[width=0.3\linewidth]{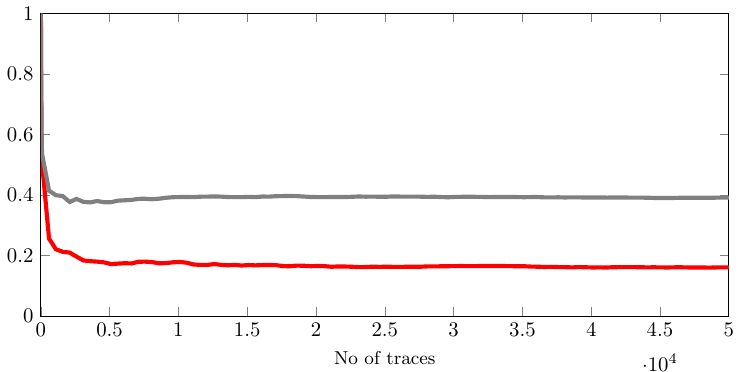} & \includegraphics[width=0.3\linewidth]{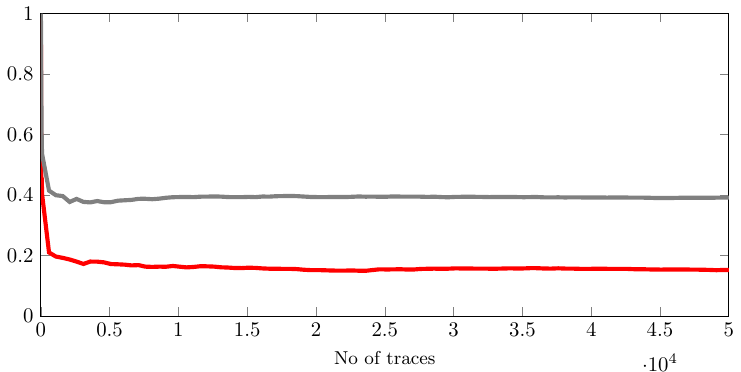} \\
        (a) sign bit & (b) exponent bits & (c) first byte of mantissa
    \end{tabular}
    \begin{tabular}{cc}
    \includegraphics[width=0.3\linewidth]{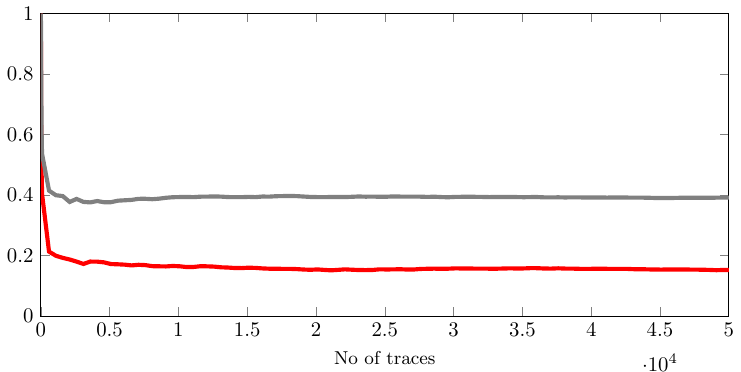} & \includegraphics[width=0.3\linewidth]{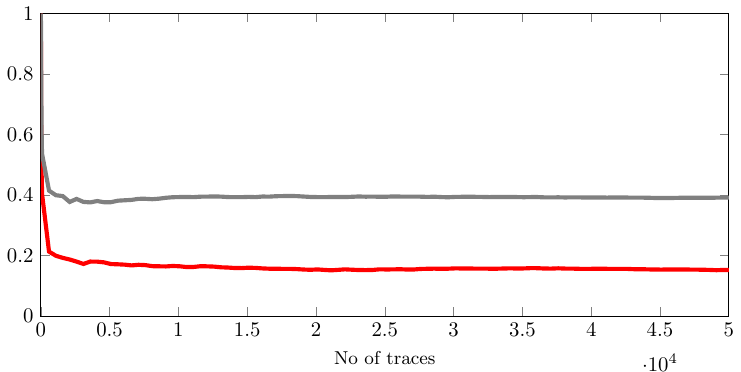} \\
       (d) second byte of mantissa & (e) last seven bits of mantissa
    \end{tabular}
    \begin{tabular}{c}
    \includegraphics[width=0.4\linewidth]{fig/_TVLSI__Shuffling_legend_traces.pdf}
    \end{tabular}
    \vspace*{-0.3cm}
    \caption{CPA attack results with different numbers of traces for the protected implementation. 
    The $y$-axis represents the absolute correlation. The red lines correspond to the correct values associated with the correct weight of $1.43$, while the gray lines correspond to incorrect values.}
    \label{fig:cpa_protected_traces}
\end{figure*}

Figures~\ref{fig:cpa_unprotected_traces} and~\ref{fig:cpa_protected_traces} illustrate the results for the unprotected and protected implementations, respectively. 
We collected a total of \(5,000\) traces for the unprotected implementation and \(50,000\) traces for the protected implementation. To evaluate the attack's effectiveness, we varied the number of traces from \(3\) to \(5,000\) for the unprotected case and up to \(50,000\) for the protected, using an appropriate step size.
For each trace count, we recorded the highest absolute correlation coefficient associated with the correct weight components and the maximum absolute correlation corresponding to incorrect weight components.

We observe that at lower trace counts, both correct and incorrect weight values exhibit high absolute correlation values in both protected and unprotected implementations. This behavior can be attributed to the limited number of data points used in the Pearson correlation calculation. Since the correlation coefficient is derived from sample means and standard deviations, small datasets result in unstable estimates, leading to significant fluctuations in correlation values. As a consequence, high correlations may occur due to random variations.
We note that a similar phenomenon has been reported in \cite{brosch2022counteract}.

As the number of traces increases, the correlation stabilizes as statistical noise averages out. The results in Figure~\ref{fig:cpa_unprotected_traces} demonstrate that in the unprotected implementation, the attack achieves success with only a few hundred traces, effectively distinguishing the correct weight components from incorrect ones. Conversely, in the protected implementation (Figure~\ref{fig:cpa_protected_traces}), even with up to \(50,000\) traces, the attack fails to differentiate correct weight components from incorrect ones. The absolute correlation values remain relatively stable, indicating that the countermeasure effectively mitigates the CPA attack.

\color{black}

\section{Discussion}
\label{sec:discuss}

\begin{figure}[tb]
    \centering
    \includegraphics[width=0.8\linewidth]{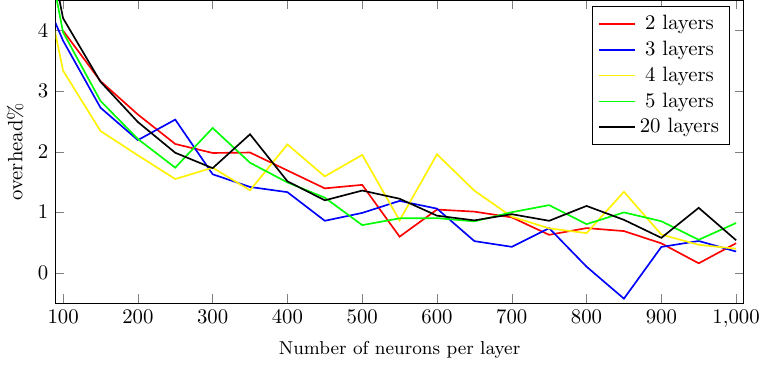}
    \caption{Computation overhead of the proposed protected neural network (MLP with varying numbers of layers) compared to standard shuffling-protected neural network implementations.
    Each layer contains the same fixed number of neurons, with ReLU as the activation function.
    The \(x-\)axis represents the fixed number of neurons per layer.}
    \label{fig:overhead}
\end{figure}

\begin{figure}[tb]
    \centering
    \includegraphics[width=0.95\linewidth]{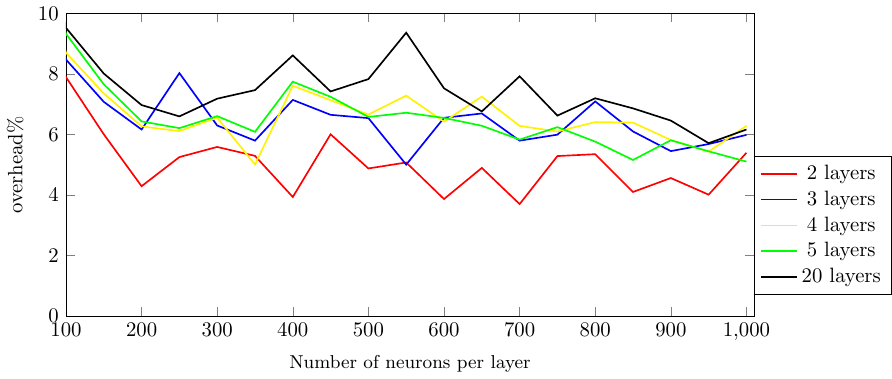}
    \caption{Computation overhead of the proposed protected neural network (MLP with varying numbers of layers) compared to unprotected implementations.
    Each layer contains the same fixed number of neurons, with ReLU as the activation function.
    The \(x-\)axis represents the fixed number of neurons per layer.}
    \label{fig:overhead_unprotected}
\end{figure}

\subsection{Overheads}
\textbf{Timing.}
Compared to the original Fisher-Yates shuffling method, the masked shuffling introduces an overhead of $3.38\times$ for $N=1000$ (i.e., with $1000$ input neurons) and $2.48\times$ for $N=100$. 

For two-layer network computation with shuffled multiplications, incorporating our proposed shuffling operation (Algorithm~\ref{alg:protected}) introduces an overhead of $4\%$ compared to the standard shuffled computation using the original Fisher-Yates algorithm (Algorithm~\ref{alg:Fisher-Yates}) when the layer has $100$ input neurons and $100$ output neurons, with ReLU as the activation function. 
When the number of input and output neurons increases to $1000$, the overhead decreases to $0.49\%$.
As the number of layers increases, the overhead remains relatively unchanged. 
Figure~\ref{fig:overhead} presents a plot depicting the overhead as a function of the number of neurons, assuming a fixed number of neurons per layer, for various layer counts in the neural network. The results indicate that the primary determining factor is the number of neurons in each layer. 
This outcome is expected, as the key variation lies in the shuffling operation preceding the multiplication, addition, and activation functions.
Given that modern network architectures often comprise several hundred neurons in each layer, we anticipate the overhead in contemporary neural networks to range between \(0.49\%-2\%\).

The decreasing trend in the overhead plot is due to the operations involved in the neuron computation: matrix multiplications and activation function calculation.
A na\"ive matrix multiplication calculation assuming the same number of neurons per layer has a time complexity of $\mathcal{O}(n^4)$, and an activation function calculation has a time complexity of $\mathcal{O}(n^2)$~\cite{orponen1994computational}.
On the other hand, the Fisher-Yates algorithm has a time complexity of $\mathcal{O}(n)$~\cite{fisher1963statistical}, and the same holds for Blakley's modular multiplication and the modulus calculation in Algorithm~\ref{alg:protected} line 4.
Therefore, it is expected that with the increasing number of neurons, the overhead of the proposed protected shuffling diminishes.

Similarly, we evaluated the computational overhead introduced by the proposed protected implementation compared to the unprotected implementation. The results, presented in Figure~\ref{fig:overhead_unprotected}, indicate a relatively stable overhead ranging between \(4\%\) and \(10\%\).

\textbf{Memory.} The memory overhead compared to Fisher-Yates shuffling-based countermeasures arises from storing the two secret arrays, \( S_1 \) and \( S_2 \). As previously noted, the size of \( S_1 \) and \( S_2 \) is determined by the maximum number of neurons in any layer of the network. Specifically, their size is equal to this maximum number minus two.

\section{Conclusion}
\label{sec:concl}
It was shown before that the shuffling-based countermeasure is effective in significantly increasing the attacker's effort required to reverse-engineer the model parameters utilizing a side-channel attack~\cite{nozaki2021shuffling,brosch2022counteract}.
However, software-based shuffling using the Fisher-Yates algorithm leaks side-channel information through its division operation, making it possible for the attacker to recover the model parameters~\cite{ganesan2023blackjack}.
In this paper, we showed how to make the algorithm resistant to that type of attack.
Our experimental results indicate that the adjusted algorithm provides the expected level of resistance while adding a small overhead that is negligible when considering the entire model computation.

For future work, it would be interesting to evaluate a combination of different hiding-based countermeasures, such as shuffling and desynchronization, or a combination of hiding and masking.
Additionally, it would be of interest to evaluate the shuffling resistance by using different preprocessing and attack analysis methods, such as Scatter~\cite{thiebeauld2018scatter}.

\appendix
In this appendix, we investigate the impact of varying input values on the performance of the attack and the performace of the attack on other weight values.

A natural question arises as to whether fixing the input values of the remaining six neurons affects the attack results. 
To address this, we randomly generated six input values and used them as fixed inputs instead of the default value of \(0.5\).
The attack was then performed on both protected and unprotected implementations.

\begin{figure*}[tb]
    \centering \small
    \begin{tabular}{ccc}
        \includegraphics[width=0.3\linewidth]{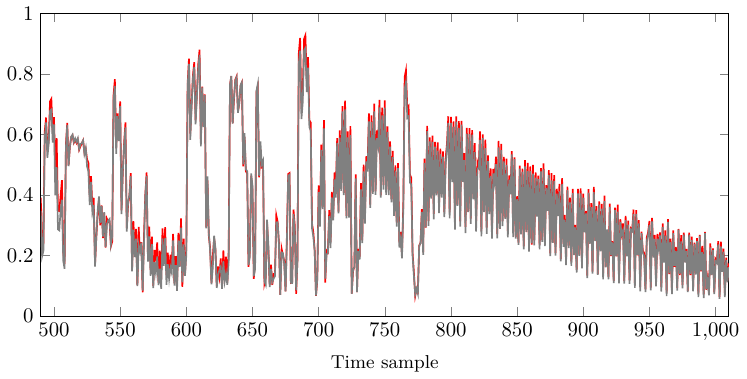} & \includegraphics[width=0.3\linewidth]{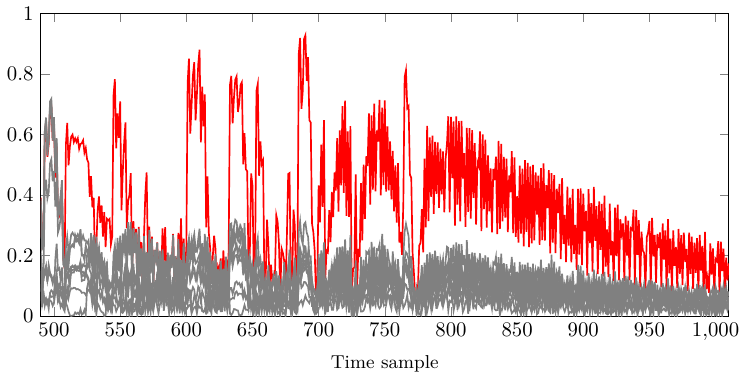} & \includegraphics[width=0.3\linewidth]{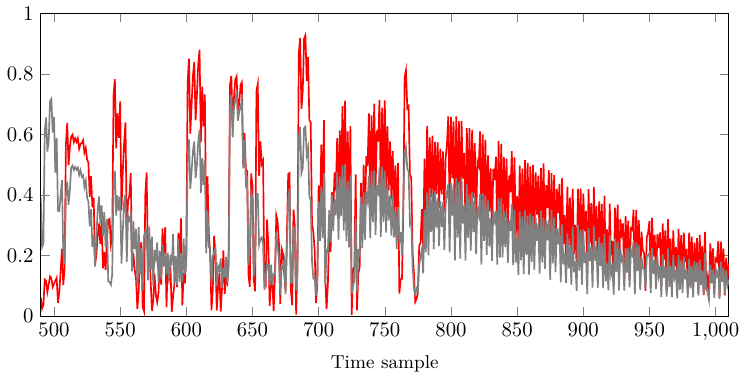} \\
        (a) sign bit & (b) exponent bits & (c) first byte of mantissa
    \end{tabular}
    \begin{tabular}{cc}
    \includegraphics[width=0.3\linewidth]{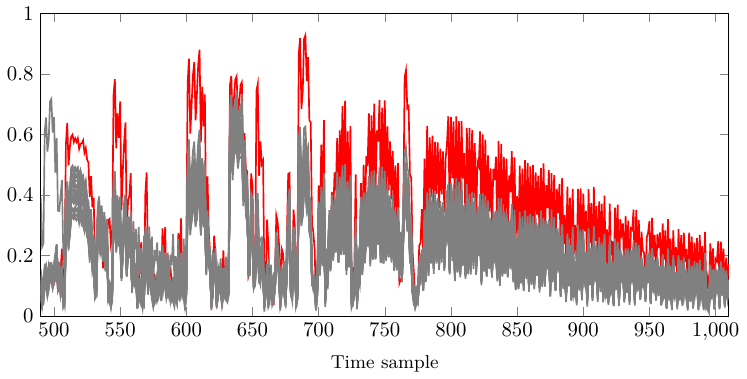} & \includegraphics[width=0.3\linewidth]{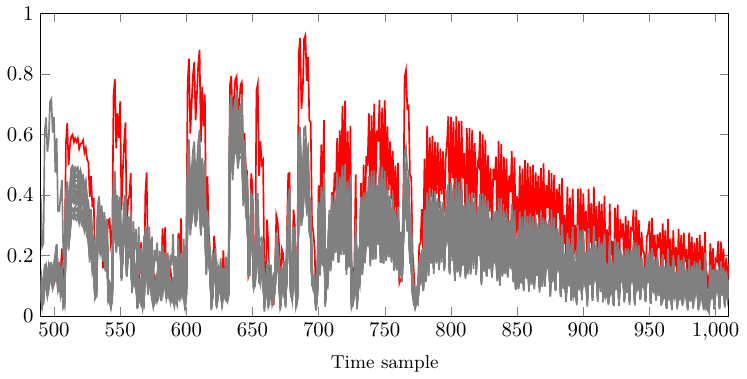} \\
       (d) second byte of mantissa & (e) last seven bits of mantissa
    \end{tabular}
    \begin{tabular}{c}
    \includegraphics[width=0.5\linewidth]{fig/legend.pdf}
    \end{tabular}
    \vspace*{-0.3cm}
    \caption{CPA attack results for the unprotected implementation, where the inputs to the six neurons (except the first) are fixed to random values.
    The $y$-axis represents the absolute correlation. 
    The red lines correspond to the correct values associated with the correct weight of $1.43$, while the gray lines correspond to incorrect values.}
    \label{fig:cpa_unprotected_difinput}
\end{figure*}

\begin{figure*}[tb]
    \centering \small
    \begin{tabular}{ccc}
        \includegraphics[width=0.3\linewidth]{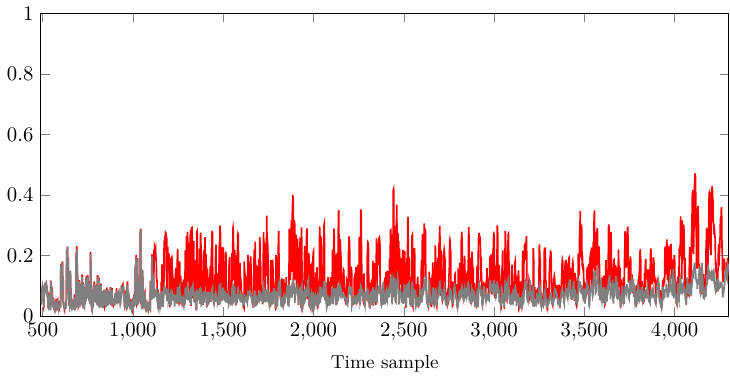} & 
        \includegraphics[width=0.3\linewidth]{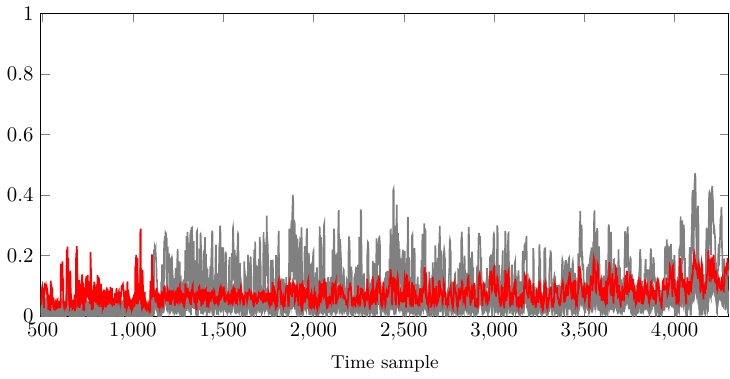} & 
        \includegraphics[width=0.3\linewidth]{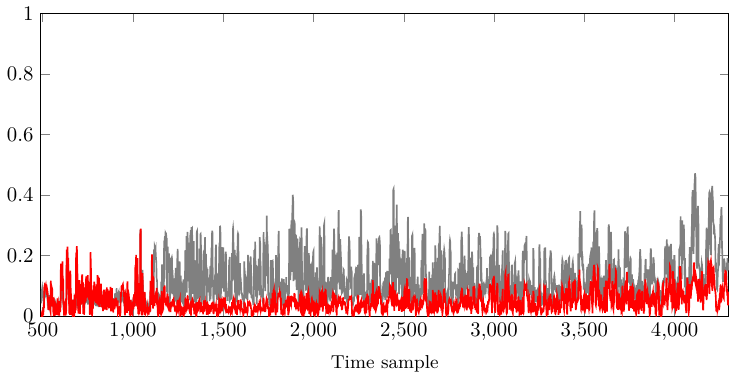} \\
        (a) sign bit & (b) exponent bits & (c) first byte of mantissa
    \end{tabular}
    \begin{tabular}{cc}
    \includegraphics[width=0.3\linewidth]{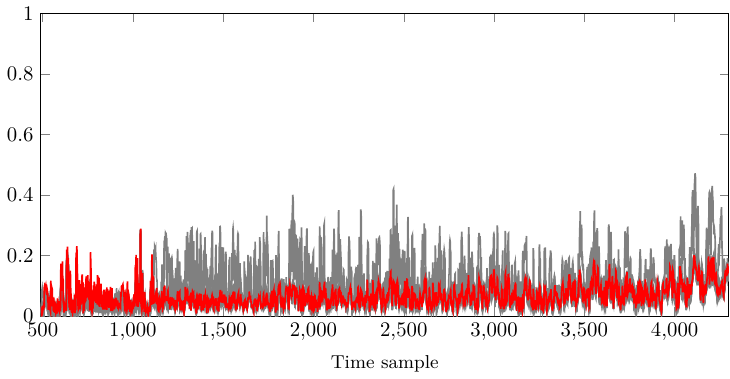} & \includegraphics[width=0.3\linewidth]{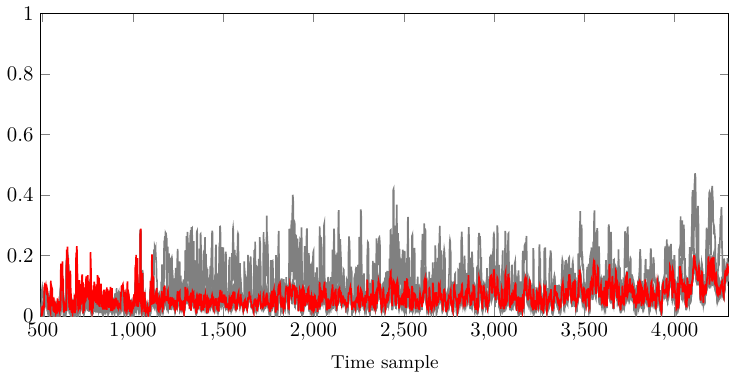} \\
       (d) second byte of mantissa & (e) last seven bits of mantissa
    \end{tabular}
    \begin{tabular}{c}
    \includegraphics[width=0.5\linewidth]{fig/legend.pdf}
    \end{tabular}
    \vspace*{-0.3cm}
    \caption{CPA attack results for the protected implementation, where the inputs to the six neurons (except the first) are fixed to random values.
    The $y$-axis represents the absolute correlation. 
    The red lines correspond to the correct values associated with the correct weight of $1.43$, while the gray lines correspond to incorrect values.}
    \label{fig:cpa_protected_difinput}
\end{figure*}

The results, presented in Figures~\ref{fig:cpa_unprotected_difinput} and~\ref{fig:cpa_protected_difinput}, correspond to the unprotected and protected implementations, respectively. 
A comparison with Figures~\ref{fig:cpa_unprotected} and~\ref{fig:cpa_protected} reveals that both the attack effectiveness and the protection measures exhibit consistent performance across varying input configurations.

\begin{figure*}[tb]
    \centering \small
    \begin{tabular}{ccc}
        \includegraphics[width=0.3\linewidth]{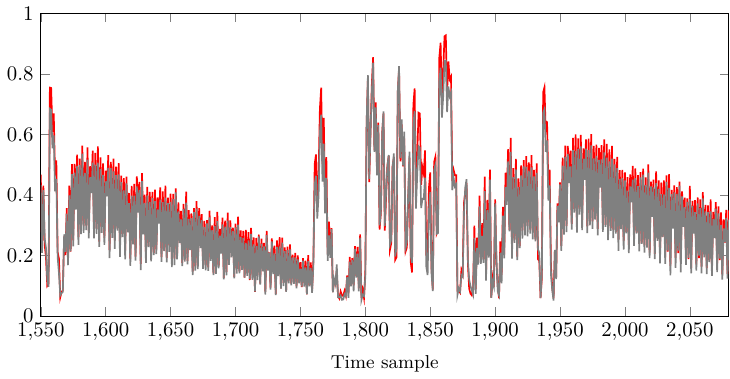} & \includegraphics[width=0.3\linewidth]{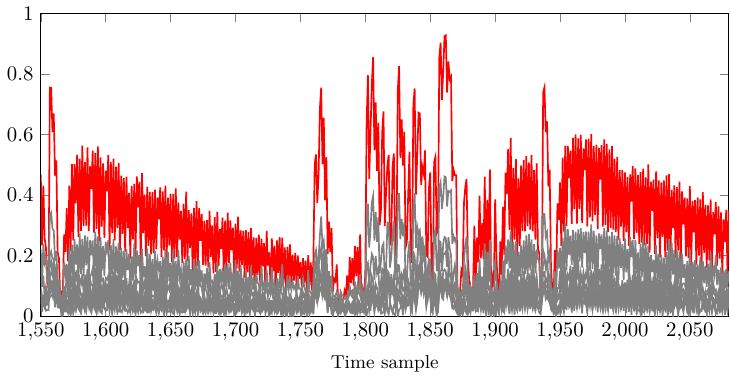} & \includegraphics[width=0.3\linewidth]{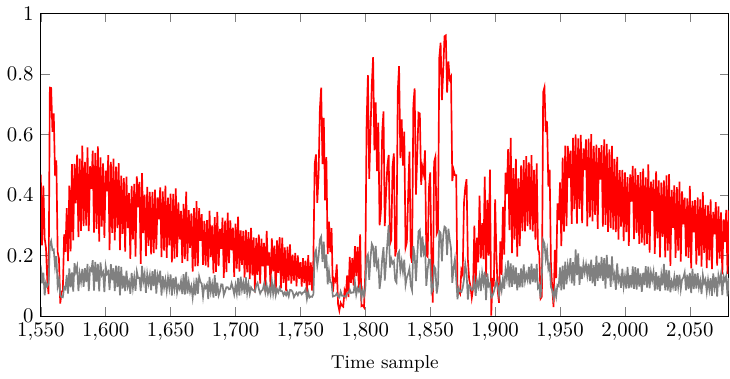} \\
        (a) sign bit & (b) exponent bits & (c) first byte of mantissa
    \end{tabular}
    \begin{tabular}{cc}
    \includegraphics[width=0.3\linewidth]{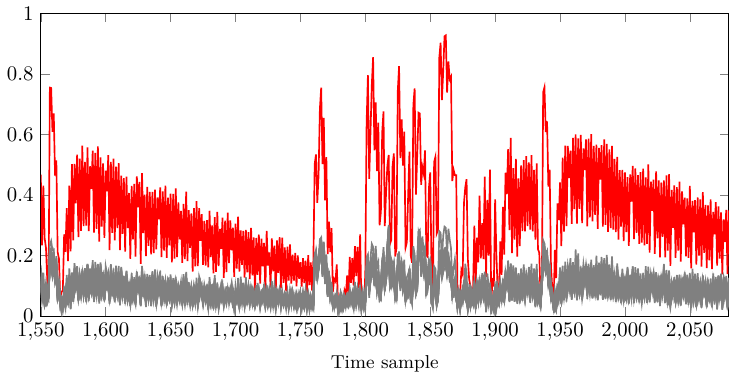} & \includegraphics[width=0.3\linewidth]{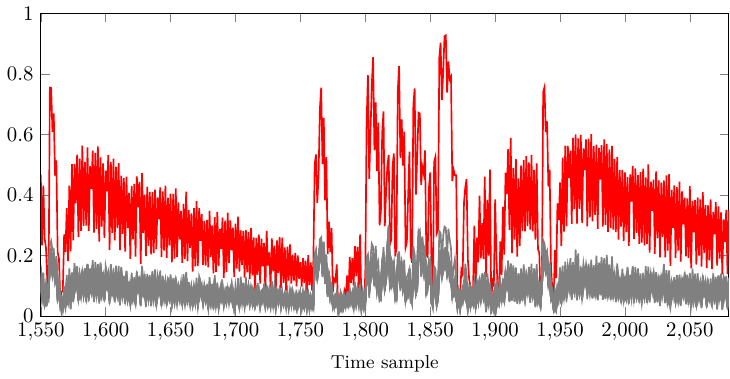} \\
       (d) second byte of mantissa & (e) last seven bits of mantissa
    \end{tabular}
    \begin{tabular}{c}
    \includegraphics[width=0.5\linewidth]{fig/legend.pdf}
    \end{tabular}
    \vspace*{-0.3cm}
    \caption{CPA attack results for recovering the weight corresponding to the third input neuron and the first output neuron of the first layer in the unprotected implementation.
    The $y$-axis represents the absolute correlation. 
    The red lines correspond to the correct values associated with the correct weight of $0.99$, while the gray lines correspond to incorrect values.}
    \label{fig:cpa_unprotected_thirdweight}
\end{figure*}

\begin{figure*}[tb]
    \centering \small
    \begin{tabular}{ccc}
        \includegraphics[width=0.3\linewidth]{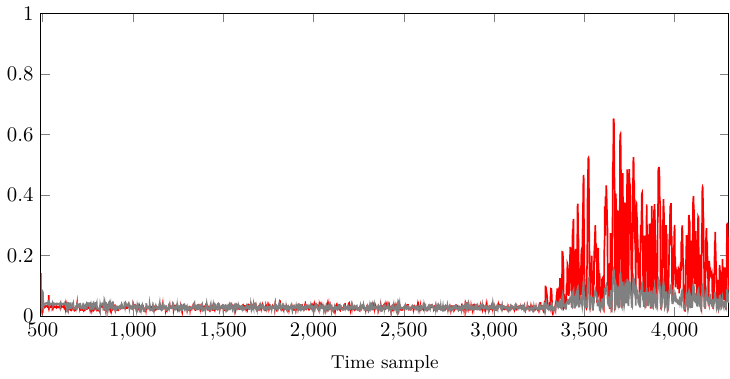} & 
        \includegraphics[width=0.3\linewidth]{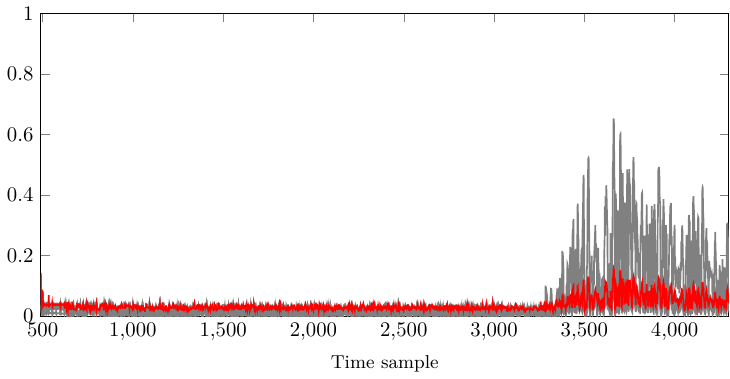} & 
        \includegraphics[width=0.3\linewidth]{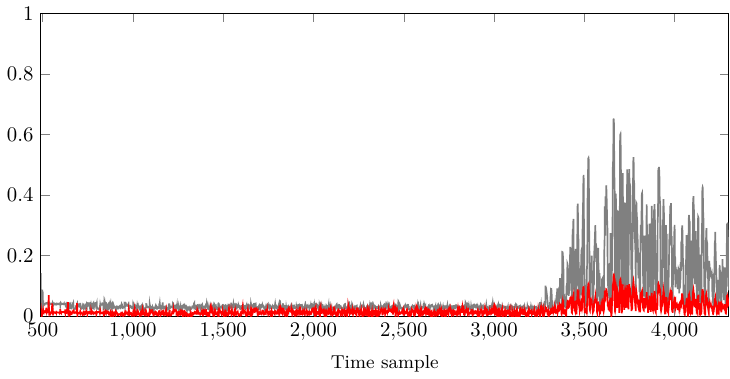} \\
        (a) sign bit & (b) exponent bits & (c) first byte of mantissa
    \end{tabular}
    \begin{tabular}{cc}
    \includegraphics[width=0.3\linewidth]{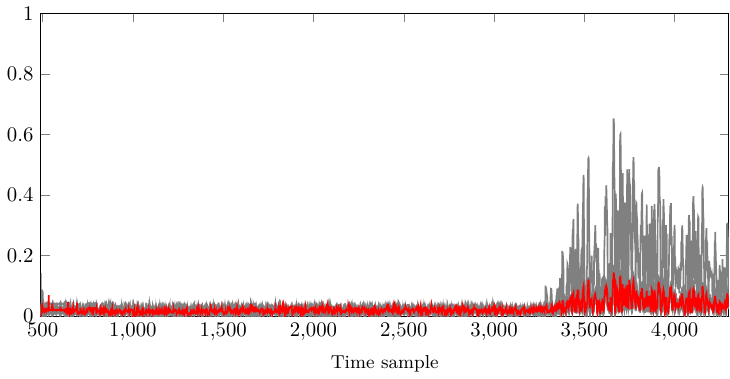} & \includegraphics[width=0.3\linewidth]{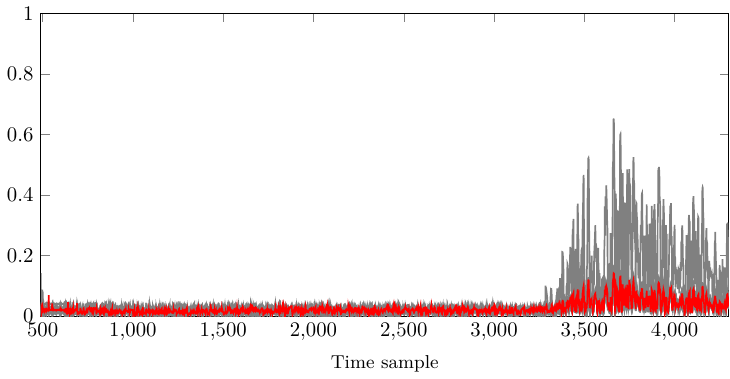} \\
       (d) second byte of mantissa & (e) last seven bits of mantissa
    \end{tabular}
    \begin{tabular}{c}
    \includegraphics[width=0.5\linewidth]{fig/legend.pdf}
    \end{tabular}
    \vspace*{-0.3cm}
    \caption{
    CPA attack results for recovering the weight corresponding to the third input neuron and the first output neuron of the first layer in the protected implementation.
    The $y$-axis represents the absolute correlation. 
    The red lines correspond to the correct values associated with the correct weight of \(0.99\), while the gray lines correspond to incorrect values.}
    \label{fig:cpa_protected_thirdweight}
\end{figure*}

To further assess the attack’s effectiveness, we attempted to recover additional weight values. 
The results, consistent for both protected and unprotected implementations, reinforce our earlier findings. 
As an example, Figures~\ref{fig:cpa_unprotected_thirdweight} and~\ref{fig:cpa_protected_thirdweight} illustrate the attack results for recovering the weight associated with the third input neuron and the first output neuron of the first layer. 
The actual weight value is \(0.99\), with the following binary representation: sign bit \(0\), exponent \(126\), first mantissa byte \(250\), second mantissa byte \(225\), and last seven bits \(36\). A comparison with Figures~\ref{fig:cpa_unprotected} and~\ref{fig:cpa_protected} further validates that the attack remains effective across different scenarios, while the countermeasure continues to provide protection in these cases.

\color{black}
\bibliographystyle{IEEEtran}
\bibliography{bibl}

\noindent
{\footnotesize \textbf{Acknowledgment.}
OpenAI’s ChatGPT-4 was used to improve the clarity and readability of this manuscript. After using this tool, the authors reviewed and edited the content as needed and take full responsibility for the manuscript's content.}

\end{document}